\documentclass[a4paper,11pt]{article}

\usepackage{combelow}
\usepackage{fancyhdr}
\usepackage[sc]{mathpazo}
\usepackage{chicago}
\usepackage{graphicx}
\usepackage{endnotes}
\usepackage{authblk}                                           
\usepackage{amssymb}
\usepackage{bbm}    
\usepackage{pgf}
\usepackage{tikz}
\usetikzlibrary{positioning,shapes,arrows,automata}
\usepackage{tensor}                                            
\usepackage{amsmath}

\usepackage{setspace}                           
\usepackage{hyperref}
\usepackage{color}
\definecolor{darkblue}{rgb}{0.0,0.0,0.3}
\hypersetup{colorlinks,breaklinks,linkcolor=black,urlcolor=black,anchorcolor=black,citecolor=black}

\usepackage{epigraph}
\usepackage[utf8x]{inputenc}
\usepackage[left=1.2in,top=1in,right=1.2in,bottom=1in,headheight=0.8in,foot=0.5in]{geometry}

\usepackage{enumitem}

\usepackage{multirow}

\usepackage{times}
\let\oldmarginpar\marginpar
\renewcommand\marginpar[1]{\oldmarginpar{\color{red}\raggedright\scriptsize #1}}

\newcommand{\mean}[1]{\ensuremath{\lf\langle #1 \rt\rangle }}

\newcommand{\comment}[1]{}
\def\lf {\ensuremath{\left}}
\def\rt {\ensuremath{\right}}

\title{{\textsc{Big Bang Singularity Resolution In Quantum Cosmology}}}

\author{}
\author{ Karim P. Y. Th\'ebault\thanks{Department of Philosophy, University of Bristol, email: \href{mailto:karim.thebault@bristol.ac.uk}{karim.thebault@bristol.ac.uk}.}}

\begin{document}

\maketitle

\begin{abstract} 
We evaluate the physical viability and logical strength of an array of putative criteria for big bang singularity resolution in quantum cosmology. Based on this analysis, we propose a mutually consistent set of constitutive conditions, which we argue should be taken to jointly define `global dynamics' and `local curvature' big bang singularity resolution in this context. Whilst the present article will focus exclusively on evaluating resolution criteria for big bang singularities in the context of finite dimensional models of quantum cosmology, it is also hoped that the core features of our analysis will be extendible to a more general analysis of criteria for quantum singularity resolution in cosmology and black hole physics.
\end{abstract}

\pagestyle{fancy}

\tableofcontents

\section{Introduction}

Classical models of the universe generically feature an initial or `big bang' singularity.\footnote{We note here that the term `big bang' is increasingly used in cosmology to refer to a hot and dense early phase of the universe as opposed to the initial singularity. Here we retain the older sense as the scope of the paper is such that there will be no ambiguity.}  That is, when we consider progressively earlier and earlier stages of the universe, observable quantities stop behaving in a physically reasonable way. A more precise mathematical characterisation of the cosmic big bang singularity can be made in terms of both a global dynamical notion of incompleteness of inextendible causal (i.e., non-spacelike) past-directed curves and a local notion of the existence of a curvature pathology \cite{penrose:1965,hawking:1970,hawking:1973,senovilla:1998,curiel:1999,curiel:2019}. That the local curvature pathology in question is physically problematic in the case of big bang singularities is unambiguously demonstrated by the existence of scalar curvature invariants that grow without bound in finite proper time along the incomplete curves in question \cite{thorpe:1977,curiel:1999}. The local curvature and global dynamics classical notions of singularity are simultaneously realised in the models of our universe that general relativity provides. Such pathological behaviour can be understood to signal the breakdown of classical cosmology at the big bang singularity and the requirement for new theoretical tools.\footnote{For critical analysis of the inevitability of the connection between singularities and the breakdown of general relativity see \cite[\S2.2]{curiel:2019} and \cite{earman:1995}. See also \cite{mattingly:2001}, \cite{crowther:2021}, and \cite{Weatherall:2022}.} 

A widespread and longstanding expectation within the physics community is that quantization of cosmological models will lead to the `avoidance' or `resolution' of singular behaviour. In contrast to the precise formal conditions for the existence of the big bang singularity, via the simultaneous realisation of the global dynamical and local curvature notions of singularity, various distinct proposals for what we mean by the resolution of singularities in quantum cosmology have been formulated in the literature. A systematic framework for the analysis of quantum singularity resolution criteria is yet to be provided. The aim of this paper is to provide such an analysis based upon an integrated analysis and evaluation of the major candidate criteria.  

Our investigation will be framed by three dimensions. The first relates to the aspect of the quantum formalism that is at the heart of the criterion at hand, in particular, whether our focus is on the observables, the quantum state, or the inner product. The second dimension relates to the aspect of the classical big bang singularity that is targeted for resolution: the local curvature pathology aspect or the global dynamics aspect. We will distinguish between local curvature and global dynamics singularity resolution criteria on the basis of classical-quantum correspondence and argue that an analysis of both aspects is required for what we mean by quantum singularity resolution to be fully understood.\footnote{See \cite[\S6.2]{Wuthrich:2006} for a related idea of `expectation-value singularity' (which would be the analogue of a local curvature quantum singularity) and `dynamical singularity' (which would be the analogue of a global dynamics quantum singularity).} The third dimension of analysis relates to the logical strength of the criterion under consideration as necessary, sufficient, or necessary and sufficient. Although we make no claim that our analysis will be exhaustive, simultaneous full exploration of these three dimensions of analysis will render it systematic and complete to the extent that we will consider virtually all candidate criteria for big bang singularity resolution that have been subject to analysis in quantum cosmology. 

As well as having as its central goal that of survey or review article, our aims in this paper will also be an evaluative one. That is, we will offer prescriptions regarding what we take to be physically well or poorly motivated conditions.  The key conclusion of our analysis will be the proposal of a mutually consistent set of \textit{constitutive conditions}, which we argue should be taken to jointly \textit{define} big bang singularity resolution in quantum cosmology. Given the role of singularity resolution in quantum cosmology as both a motivation for and indicator of success in quantum gravity research, the abiding value of our analysis for researchers in quantum gravity should hopefully be clear. 

The constitutive conditions are broken-down as follows. First, with regard to global dynamics singularity resolution, we will endorse unitarity as a necessary but not sufficient condition. The strongest candidate for a necessary and sufficient condition will then be isolated as the condition of quantum hyperbolicity \cite{bojowald:2007}. However, the applicability of this criterion is limited by the calculation demands of its unambiguous establishment. In the context of local curvature  singularity resolution, we will endorse two constitutive criteria, one sufficient condition and one necessary condition. The sufficient condition we will identify is the boundedness of self-adjoint operators on the physical Hilbert space whose classical correlates are divergent \cite{ashtekar:2011}. The necessary condition is finiteness of expectation values  \cite{brunnemann:2006,Gryb:2017a}.  A summary of our proposed constitutive  conditions sorted along the three lines of analysis is provided in Table \ref{table1}.

A further value of our analysis comes from exploration of secondary, \textit{indicative conditions}. Such conditions will prove of considerable diagnostic value due to the calculation complexity of establishing the full set of constitutive conditions. In this sense our analysis can be expected to be of specific heuristic value to researchers in quantum cosmology. We will categorize as an indicative sufficient condition for global dynamics singularity resolution the failure of the expectation value to reach zero for observables that are classicaly zero in the singular regime \cite{lund:1973,gotay:1980,gotay:1983}. On a formally related basis, we will endorse as a necessary and sufficient indicative condition for both global dynamics and local curvature singularity resolution the existence of self-adjoint `slow' clocks  \cite{gotay:1983,gielen:2022}. Finally, we will also endorse the vanishing of the wavefunction on three-geometries of zero volume as a plausible sufficient indicative condition for global dynamics singularity resolution.
   
The framework described is both tentative and limited in range of application. It is hoped, however, that various aspects of our analysis will be extendible to a more general analysis of quantum singularity resolution including, in particular, black hole curvature singularities and the more general class of physically realistic models with cosmological singularities. Such more general analysis of singularity resolution is left for future work.

\begin{center}
\begin{table}[]
\centering
\caption{Proposed constitutive conditions for quantum singularity resolution.}
\bigskip
\label{table1}
\begin{tabular}{c|c|c|ll}
\cline{2-3}
\multicolumn{1}{l|}{}                & \multicolumn{1}{|c|}{\textbf{Local Curvature}} & \multicolumn{1}{|c|}{\textbf{Global Dynamics}} &  &  \\ \cline{1-3}
\multicolumn{1}{|c|}{\textbf{Necessary}} &       Finite Expectation Values                   &       Unitarity                     &  &  \\ \cline{1-3}
\multicolumn{1}{|c|}{\textbf{Sufficient}}   &   \shortstack{Bounded Operators }                 &                            &  &  \\ \cline{1-3}
\multicolumn{1}{|c|}{\textbf{Necessary \& Sufficient}}        &                      &    Quantum Hyperbolicity                       &  &  \\ \cline{1-3}
\end{tabular}
\end{table}   
\end{center}


\section{Methodological Principles}

Even the restricted context of cosmology and big bang singularities proves to provide a field of candidate conditions too large to plausibly survey within a single article. Moreover, our ability to evaluate such conditions is severely limited by the lack of a formally and empirically well-established theory of quantum gravity and, furthermore, the intractability of anything but the simplest quantum cosmological models. In order to render our analysis well-constrained and tractable we will adopt three methodological principles: \textit{Model Based Reasoning}, \textit{Classical Correspondence}, and \textit{Interpretational Neutrality}. These principles are both prima facie plausible and consistent with mainstream practice within the community of researchers in contemporary quantum cosmology. As such they are expected to be largely uncontroversial, although certainly not incontestable. 

\subsection{Model-Based Reasoning}

The principal challenges in evaluating candidate conditions for singularity resolution in quantum cosmology are the lack of a formally and empirically well-established theory of quantum gravity and the tractability limitations of all but the most simple models. On the first point the contrast with general relativity is, of course, clear. In that context, we do have access to theory that is formally and empirically well-established in that it admits a fully rigorous mathematical statement and has been subject to stringent experimental and observational testing. The issue of tractability is less clear cut in the classical context, however. In particular, whilst there are fully tractable analytic solutions in the homogenous and isotropic context of the Friedmann–Lema\^{\i}tre–Robertson–Walker (FLRW) models, the same is not true for the general class of realistic cosmological models which must be studied perturbatively.

Crucially, the arguments for the generic existence of big bang singularities within classical cosmology do not depend upon explicit calculation within highly symmetric models. Rather, the logic of the Penrose-Hawking singularity theorems relies upon proof under general conditions \cite{hawking:1973}. The study of singularities in classical cosmology therefore has the attractive feature that even though we cannot analytically probe the full structure of inhomogeneous and anisotropic cosmological models space, it is possible to provide compelling physical arguments that such cosmologies will be singular whenever they satisfy the antecedents of Penrose-Hawking type singularity theorems. It has thus been possible to establish the generic existence of the classical big bang cosmological singularity via strong theoretical arguments that operate at the level of the full solution space of the theory.

The style of reasoning deployed in the context of analysis of big bang singularity resolution in quantum cosmology is by necessity rather different. There are, at present, no singularity theorems valid for the full domain of quantum cosmology.\footnote{The results that exist are formulated within the semi-classical regime. See for example \cite{wall:2013a,fewster:2022}. The connection between these results and the resolution proposals is that we will discuss here in the context of the mini-superspace framework for quantum cosmology is an interesting topic that warrants further study.} Moreover, lack of empirical data and fully reliable perturbative techniques means that a highly limited range of reliable constraints in model building are available. Finally, the analytically tractable regime in quantum cosmological models is even more restricted than in the classical context. Most extant treatments are focused on the quantization of finite dimensional truncations of classical cosmological models, for the most part either homogenous and isotropic mini-superspace models or homogenous but anisotropic Bianchi models.\footnote{We should note here that there do exist a limited set of singularity resolution results within field theoretic approaches to quantum gravity and quantum cosmology, in particular in the context of symmetry reduced but infinite dimensional midi-superspace models. See  \cite{barbero:2010} and \cite[\S6]{ashtekar:2011} for discussion in the context of metric and loop approaches and   \cite{calcagni:2014,oriti:2017,Cesare:2017,marchetti:2021} for group field theory approaches. There are also interesting results in so-called Gowdy models that feature gravitational waves. See \shortcite{tarrio:2013,deBlas:2017}.
Such results are of course still model-based as a degree of idealisation is necessary to extract solvable dynamical equations. Furthermore, we would expect most, if not all, of the main ideas relating to singularity resolution discussed here to be applicable to the field theoretic domain, although this extension warrants its own detailed treatment.} Even in that context it is often necessary to apply numerical techniques to explore the structure of the relevant solutions. 

These worries notwithstanding, one can certainly ask, and in some cases answer,  physically interesting questions regarding the resolution or not of singularities within this restricted context. The approach that is followed within the literature is a form of \textit{model-based reasoning} in which \textit{idealised models} mediate our understanding of the properties of full, physically representational, theory \cite{sep-models-science}. The reliability of such reasoning is highly sensitive to the \textit{stability} of the salient features under de-idealization to a fuller and more physically realistic model class. The same idea is expressed by \citeN{ashtekar:2011} in terms of the quantum theory of the idealised system `capturing the relevant qualitative features' of the quantum theory of the de-idealised system (p. 213001).   

In this context it will prove instructive to differentiate two importantly distinct questions of stability in the context of cosmological singularity resolution criteria:
\begin{itemize}
\item [] \textbf{Stability of Result}: Given some criterion for singularity resolution $C$, how stable is the result of application of this criterion under relaxation of relevant assumptions such as symmetry. i.e. do we have good reason to believe the result of resolution (or not) under the criterion depends upon unrealistic physical features of the model. 
\item [] \textbf{Stability of Applicability}: Given some criterion for singularity resolution $C$, how stable is applicability of this criterion under relaxation of relevant assumptions such as symmetry. i.e. do we have good reason to believe the application of the criterion depends upon unrealistic physical features of the model.   
\end{itemize}
Since the focus of our investigation is the analysis of criteria for singularity resolution, it is stability of the second rather than first sense that is important. That is, we will not demand that applying the same criterion to relevantly finite and infinite dimensional models should produce stable results with regard to singularity resolution.\footnote{For discussions  regarding stability of within and between the relevant model classes see \cite{kuchavr:1989,sinha:1991,ishikawa:1993,ashtekar:2009}. For work on the relationship between symmetry reduced Loop Quantum Cosmology and the full theory of Loop Quantum Gravity see \cite{assanioussi:2018,olmedo:2019}. } Rather, what is important in our approach is that any criterion considered must be such that it can be determinately applied to both simple toy models of quantum cosmology (e.g. finite dimensional models) and de-idealised realistic models (e.g. models with infinite degrees of freedom). A criterion will fail stability of applicability when it proves to be only applicable to simple models. As noted, a criterion can satistify stability of applicability even when it does not show stability of result.   

In the context of stability of result it is worth making a short digression regarding the interesting connection between stability of the results of resolution criteria and the Belinsky-Khalatnikov-Lifshitz (BKL) conjecture. The BKL conjecture is a classical hypothesis of the existence of dynamical behaviour such that at the asymptotic approach to the singularity there is a decoupling of spatial points leading to a scenario in which the universe  exhibits the dynamics of a finite dimensional Bianchi model at each spatial point \cite{belinskii:1982}. It has been proved that a dense set of inhomogeneous GR solutions obey the BKL conjecture \cite{andersson:2001}.

The principal physical mechanism behind BKL behaviour is that the dynamical influence of spatial derivatives embodying communication between spatial points is overwhelmed by the time dependence of the local dynamics as the singularity is approached. The generic existence of such behaviour can be supported by evidence based on numerical investigations \cite{berger:2002,garfinkle:2007}. In the context of classical Bianchi IX models it has been argued that non-singular classical behaviour demonstrated in relational variables  \cite{koslowski:2018} should be stable under de-idealisation to the full classical theory on the basis of a proof that spatial points decouple in the approach to the singularity and evolve as independent Bianchi IX systems \cite{andersson:2001}, combined with the assumption of a `stiff fluid' component of matter (i.e., a perfect fluid with pressure equal to energy density).

The relevance of BKL to stability of result of resolution criteria can then be understood in terms of the potential for a 
quantum version of the BKL conjecture to secure the robustness of singularity resolution under de-idealisation from finite to infinite dimensions. In particular, the idea would be that lessons regarding the quantum nature of the big-bang singularities in Bianchi models may be valid much more generally precisely because of the decoupling between spatial points in the BKL regime \cite{ashtekar:2009}. A BKL-based argument for the stability of resolution results under de-ideasliation away from quantum Bianchi models is limited, however, by the fact that quantum bouncing models can generally be expected to prevent the asymptotic approach to singularity and thus resolution would happen away from the limit in which the BKL conjecture would be valid \cite{bojowald:2011c,bojowald:2020}. In such circumstances the singularity \textit{would} still be avoided by a quantum bounce, however, there would no longer by an argument available that this dynamics could be analysed via Bianchi models based on the BKL conjecture. 

Putting these fascinating issues to one-side, as already noted, the demand that is crucial to our analysis relates to stability of applicability. That is, the \textit{applicablity} of any viable criterion for singularity resolution in quantum cosmology should not be unstable under de-idealisation outside the highly symmetric sector. Criteria for resolution should in principle be applicable to a full theory of quantum cosmology.\footnote{This aspect of our approach generalises the line of reasoning adopted by \cite{Warrier:2022} in the specific context of the idea of the wavefunction vanishing as a putative singularity resolution criterion.} We will further limit ourselves by focusing upon  minimally speculative models for quantum gravity based upon quantization of general relativity following standard canonical methods (including loop based quantization). Thus, we will not consider stability in the broader sense of applicability to different conceptualisation of what a full theory of quantum cosmology might be. That said, all the candidate conditions we will consider rely on core formal structures of quantum theory. They should be expected to be stable in this sense, at least so long as the relevant theory of quantum cosmology is \text{interpretationallly neutral} in the sense that we will consider shortly.
 
In this context, in order to render our evaluative task a tractable one, a further restriction will be placed upon the formalisation of models of quantum gravity that we will consider. This is in terms of a restriction to formulations of quantum cosmological models which are formulated based upon Hilbert spaces and operators. Primarily these will be considered in the context of canonical quantizations of classical cosmological models. We will thus not include in our analysis approaches to singularity resolution based upon gravitational path integrals or holography, or, as would be relevant to that context, formalisation of resolution criteria in terms of behaviour of correlation functions.\footnote{An interesting example of recent work in this direction, which is in fact inspired by loop based resolution proposals, is \shortcite{bodendorfer:2019,bodendorfer:2019b}. } Furthermore, and relatedly, the present analysis will also not include discussion of proposals for singularity resolution based upon signature change described in terms of (Euclidean) path integrals along the lines of the Hartle-Hawking no boundary proposal \cite{hartle:1983,vilenkin:1984}. Extension of the framework developed here to such contexts would be an interesting and valuable exercise for future work.\footnote{Recent work due to \citeN{brahma:2018} suggests that the combination of a Hartle-Hawking type no-boundary proposal with the loop quantum cosmology framework, leads to an interesting interplay between singularity resolution, smooth initial conditions and inflation.}   

Finally, it is worth pointing to a further particular value of the model-based approach: since we are evaluating resolution criteria and the symmetry reduced models in questions are, strictly speaking, `models of the theory' in the sense that they are quantizations of Einstein spacetimes, their investigation will be sufficient to establish counter-examples. That is, given that we can motivate plausible resolution conditions on a physical basis, we can then seek to establish counter-examples based upon the symmetry reduced models and numerical results where available. What we shall find in a number of cases is that prime facie plausible necessary or sufficient criteria for singularity resolution can be rejected on the basis of the study of simple models. Whilst a model-based analysis is of course not suited to demonstrating the general reliability or applicability of a candidate condition, it does provide a strong basis to demonstrate by counter-example unreliability and inapplicability.

\subsection{Classical Correspondence}

An important implicit assumption in our analysis thus far has been that we can make sense of what it means for a condition for singularity resolution to fail. For this to be the case we must assume that there is some reasonable physical basis to understand what it means for a criterion of resolution to obtain in a given model, but for the resolution itself to be taken to \textit{not} to have taken place. The crucial physical resource we are able to rely upon here is correspondence to the two classical notions of singularity that jointly apply to the big bang singularity in FLRW models.\footnote{It is in this context that the restricted focus of our analysis is particular helpful. The full taxonomy of types of classical singularity recognised in the contemporary literature include exotic and varied conceptualisations of what we might include under the label of `spacetime singularity' in which the local curvature and global dynamics senses come apart. Examples include boundary constructions \shortcite{hawking:1967,geroch:1968,schmidt:1971,geroch:1982,curiel:1999}, sudden singularities \cite{barrow:2004,barrow:2004b,nojiri:2004,cattoen:2005,cotsakis:2005,fernandez:2014,beltran:2016,fernandez:2004}, big rip singularities \cite{caldwell:2002,caldwell:2003,chimento:2004,elizalde:2004,nojiri:2005,dabrowski:2006,fernandez:2014,harada:2018} and failure of extendibility and regularity \cite{fournodavlos:2020,sbierski:2021}. The project of providing a systematic and unified categorisation of what it means for a classical spacetime to be singular is thus a daunting one, if not entirely hopeless one \cite{curiel:1999,curiel:2019}. } That is, we will focus on correspondence via the semi-classical limit to the key physical characteristics of the classical big bang singularity in terms of the combination of incompleteness of inextendible causal past-directed curves and the existence of a curvature pathology. The big bang singularity simultaneously corresponds to the incompleteness of \textit{all} causal past-directed curves\footnote{There is a wider range of (global) big bang singularity types in FLRW cosmology given the relaxation of energy conditions. See \cite{harada:2018} for a classification scheme.}  within an inextendible spacetime and the existence of a curvature pathology as the volume of the universe tends to zero.\footnote{It should be noted that that classical correspondence according to our characterisation is a fairly conservative principle in the sense that it rests on the assumption of general relativity as the appropriate theory of \textit{classical} cosmology.  Classical modified gravity scenarios, such as higher-derivative theories or scalar-tensor gravity, could plausibly lead to different realisation of the big bang singularity, or even classical singularity resolution, and thus to more complex scenario for the evaluation of quantum singularity resolution criteria. For a review of cosmology and modified gravity which includes discussion of singularities see \shortcite{clifton:2012}
Thanks to an anonymous referee for suggesting this issue.}  

Assuming that the cosmological model we are considering is globally hyperbolic and diffeomorphic to $\mathbb{R} \times \sigma$, with $\sigma$ a Riemannian three-geometry, we can then understand the global singularity in terms of failure of the Einstein equations to be well-posed as a system of PDEs for all values $t \in\mathbb{R}$ \cite{brunnemann:2006}.\footnote{An attractive consequence of framing the global condition in this manner, rather than focusing on geodesic incompleteness, is that there is a more clear connection then obtains between the relevant global dynamical singular behaviour and the conception of a singularity as expressed in terms of breakdown in regularity \cite{fournodavlos:2020,sbierski:2021}.} The semi-classical analogue of the global dynamics singularity, characterised as a failure of well-posedness of the classical equations of motion, can then be understood simply via the breakdown of the semi-classical equations of motion, which can be characterised, for example, via the moment expansion of the full set of observables. The local curvature pathology singularity can be understood in terms of divergence of polynomial scalars built out of the Riemann tensor and the metric.  Since the curvature scalars in cosmology can be characterised via powers of the inverse scale factor we can then unambiguously characterise the curvature pathology associated with the big bang singularity via the behaviour of the inverse scale factor. The semi-classical analogue of the local curvature big bang singularity can thus be characterised via the divergent behaviour of the scale factor in the semi-classical regime.

It is important to note that there are good reasons to take the global dynamics and local curvature notions of a classical big bang singularity as interdependent. In particular, although the divergence of scalar polynomials in the metric and the Riemann tensor is frame independent, the relevant unboundedly large values may occur only at infinity for a given casual curve \cite{joshi:2014}. Observers on such a curve would experience no physical pathology. For this reason, the classical notion of curvature singularity is best defined in terms the existence of scalar curvature invariants that grow without bound \textit{in finite proper time} along incomplete of inextendible causal past-directed curves \cite{thorpe:1977}. The classical local curvature sense of big bang singularity thus needs to be connected to the global dynamics sense to be interpreted as an unambiguous physical pathology. This might be understood to suggest that correspondence to the global dynamics sense of singularity should be the primary focus of singularity resolution criteria. However, although the interdependence is less direct, there are good reasons to take the physically reliable application of the global dynamics sense to require reference to the local curvature sense. This is particularly true in the case of the big bang singularity due to the fact that the classical singularity is an example of a \textit{strong} curvature singularity in which particular integrals of curvature invariants diverge as a critical value of the affine parameter is approached and any physical detector is inevitably destroyed \cite{ellis:1977,tipler:1977,krolak:1986}. Strong singularities have been conjectured to generically be associated with geodesic incompleteness \cite{krolak:1987,tipler:1980} and thus link the local curvature and global dynamics senses of singularity. To understand the nature of a classical singularity one needs to analyse its different properties in unison \cite{singh:2012} and thus there is a good motivation for us to look for criteria for singularity resolution that have suitable correspondence to both the local curvature and global dynamics sense of classical singularity. 

The implications of classical correspondence for the \textit{failure} of candidate necessary and/or sufficient criteria for local curvature (global dynamics) singularity in quantum cosmology resolution are then as follows: Classical correspondence implies that a local curvature (global dynamics) \textit{necessary} condition \textit{fails} when there exists models for which the classical local curvature (global dynamics) big bang singularity exists, the condition does not obtain in the quantum theory, and yet the correspondence between the classical and quantum singular behaviours breaks down \textit{in some physically relevant sense} such that the singularity is resolved. Correspondingly, a local curvature (global dynamics) \textit{sufficient} condition \textit{fails} when when there exists models for which the classical (local curvature or global dynamics) singularity exists, the condition obtains in the quantum theory, and yet the correspondence between the classical and quantum singular behaviours holds \textit{in some physically relevant sense} such that the singularity is not resolved. Clearly much will depend upon how we understand the 'physically relevant sense'.  Let us illustrate what we mean by reference to the discussion of \S \ref{unitarity}. There we will argue that the criterion of essential self-adjointness of the Hamiltonian can be understood to fail as a necessary condition for global dynamics singularity resolution on the basis of: i) the assumption that the physically relevant sense of global dynamics singularity resolution is having well-defined equations of motion at all times; and ii) reference to the variety of models which have determinate unitary quantum evolution parametrised by a family of non-unique self-adjoint extensions, despite the failure of classical determinate evolution per geodesic incompleteness. 

Ultimately, reliable criteria for singularity resolution should be robust under different physically relevant choices of how to best understand the classical-quantum correspondence and we should also expect the relevant global dynamics or local curvature notions to be at least consistent and ideally mutually reinforcing. For this reason we shall treat classical correspondence between local curvature and global dynamics criteria as a basic desiderata. That is, we will argue against, for example, global dynamics criteria for singularity resolution whose satisfaction is consistent with the failure of physically reasonable local curvature criteria. Thus classical correspondence motivates us ultimately to look for quantum singularity resolution criteria that cut across the local curvature/global dynamics divide. 

\subsection{Interpretational Neutrality} 

Two serious interpretational problems stalk any attempt to analyse quantum cosmological models. The first is the \textit{cosmological quantum measurement problem}. This is the acute version of the general measurement problem that obtains in the context of quantum cosmology. Foundational questions regarding the interpretation of quantum mechanics become more pressing at the cosmological scale \cite{Bell:2001}. Indeed, it was the specific problem of interpreting quantum cosmology that spurred the original development of the Everett interpretation \cite{everett:2012}. Two specific and interconnected aspects of the cosmological interpretation problem relate to the interpretation of probability and the wavefunction.\footnote{The first issue might seem to put constraints on the second in that any interpretation that is predicated on a frequentist account of probability that requires an ensemble of quantum systems might plausibly be taken to be ruled as inapplicable to quantum cosmology \cite{evans:2016}. However, even in this regard there is no clear consensus \cite{vilenkin:1989}.} 

The question of the interpretation of the wavefunction, although unresolved, does at least afford us an attractive means to differentiate interpretational options. In particular, whilst one group of interpretations take the wavefunction to refer to the physical state of some quantum system \cite{Bohm,everett:1957}, another take the wavefunction to refer to knowledge or information \cite{heisenberg:1958,peierls:1991,caves:2002,fuchs:2002,HarriganSpekkens10}. This division between so-called `$\psi$-ontic' and `$\psi$-epistemic' interpretations can then be supplemented by a parallel distinction between $\psi$-complete and $\psi$-incomplete interpretations where incompleteness indicates the existence of supplementary `hidden variables' in addition to the quantum state.\footnote{See \cite{hance:2021} for an argument that the ontic/epistemic distinction may not be an exclusive one.} Significantly, although various important no-go results have been derived the field of candidate interpretations remains almost unconstrained and debate regarding the viability of the various options remains, to a large degree, deadlocked.\footnote{See for example \cite{ben:2017}.}

What is significant for our current project is the role that interpretations should or should not play in the articulation and evaluation of conditions for singularity resolution. Given the deadlocked debate, we will take a stance of censorious neutrality. That is, we will not only refrain from adopting any particular interpretation within our investigation, but rule out as viable candidates conditions that rely on a particular interpretational approach for their application.\footnote{This is very much in line with remarks regarding the strength of loop quantum cosmology approaches to singularity resolution made in \cite[\S7.3.1]{bojowald:2011c}.} Since, in our view, the debate regarding the interpretation of quantum mechanics is as yet entirely unresolved, we adopt a stance of interpretational neutrality vis-\`a-vis the measurement problem, as well as the connected issues of interpretation of probability and the wavefunction, as a methodological prerequisite for evaluation of conditions for singularity resolution in quantum cosmology. Our adoption of this stance should be understood in broadly pragmatic terms. It rests upon the fact that after more that a hundred years of pursuit, our search for means of identification of the `true' interpretation of quantum theory has so far proved to be a vain one.\footnote{One implication of the methodological principle of interpretational neutrality is that our investigation is delimited to exclude the family of conditions for singularity resolution that are specifically founded upon the de Broglie-Bohm type interpretations. In particular, we will not consider conceptions of singularity resolution that are either specifically founded upon the behaviour of Bohmian trajectories \cite{falciano:2015} or whose physical interpretation is largely dependent upon the Bohmian approach \cite{demaerel:2019}. This delimitation of our enquiry is principally motivated by limitations of space and we take it that a detailed comparative investigation into the connections between Bohmian-focused resolution criteria and those discussed in this paper would be a very worthwhile project.}

The second serious interpretational problem that besets the study of quantum cosmological models is the  \textit{problem of time}. What is called \textit{the} problem of time is in reality a cluster of interconnected problems \cite{Isham:1992,Kuchar:1991,anderson:2012b,anderson:2017}. Some of these issues are more formal and some more interpretational. What is most relevant here is the question of whether the evolution of the quantum observables should be understood as an intrinsic or extrinsic notion in systems where the classical Hamiltonian is constrained to be zero. 

The orthodox position is to take an \textit{intrinsic time} approach in which the observables are understood to evolve according to a non-unique and typically time-dependent Hamiltonian on the physical Hilbert space as dictated by the choice of a physical variable as an internal clock parameter.\footnote{Such approaches are variously described as `evolving constants of the motion' or `complete observables' and share a common conceptual core with differing technical applications. See \cite{Page:1983,Rovelli:1990,Rovelli:1991,Rovelli:2002,gambini:2001,Dittrich:2007,Dittrich:2006,gambini:2009}.} In the context of quantum cosmology intrinsic time approaches correspond to de-parametrizations of a fundamentally timeless Wheeler-DeWitt type equation with respect to a physical variable. In symmetry reduced minisuperspace models the clock variable is typically selected as either the scalar field or some function of the scale factor.  

An alternative, less orthodox, position is to take an \textit{extrinsic time} approach in which evolution is understood as an independent feature of the dynamics even in systems where the classical Hamiltonian is constrained to be zero.\footnote{Extrinsic time approaches can be justified on the basis of the observation that the integral curves of the vector field generated by the Hamiltonian constraint in globally reparametrization invariant theories \textit{should not} be understood as representing equivalence classes of physically indistinguishable states since the standard Dirac analysis does not apply to these models \cite{Barbour:2008,Pons:2005,Pons:2010}. On this view, successive points along a particular integral curve should be taken to represent physically distinct physical states  \cite{gryb:2011,gryb:2014,Gryb:2016a} and the vanishing of the Hamiltonian does not mean that time evolution must be, rather paradoxically, classified as the `unfolding of a gauge transformation' \cite{Pons:2005}. Further support for this viewpoint can be found in the excellent recent analysis of \cite{Landsman:2021}. See, in particular, Theorem 8.19.} An extrinsic time approach implies that observable operators evolve with respect to an unobservable time label whose role in the formalism is to distinguish distinct successive states of the universe. In the context of quantum cosmology such approaches correspond to a unimodular (or fluid) time approach in which extrinsic quantum evolution is defined relative to a parameter that is canonically conjugate to the cosmological constant or perfect fluid \cite{unruh:1989,gielen:2016,Gryb:2017a,Gryb:2017b,Gryb:2018}. 

We will again adopt a stance of interpretational neutrality within our analysis such that we will assume there to be a methodological prerequisite in the evaluation of conditions for singularity resolution in quantum cosmology under which such conditions should be applicable to both intrinsic and extrinsic time approaches. This feature is particularly important since recent analysis by Gielen and Men\'{e}ndez-Pidal \citeyear{gielen:2020,gielen:2022} has demonstrated that a single symmetry reduced model can be studied simultaneously within the intrinsic and extrinsic time approaches and that the applicability of a single commensurable conception of singularity resolution can thus prove a plausible evaluative criterion with regard to the interpretative question of whether time should be understood as intrinsic or extrinsic.\footnote{In this context it is worth mentioning the connected idea of \textit{clock neutrality} in terms of the independence of the dynamics from choice of intrinsic time \cite{hohn:2019,hohn:2020a}. What is important for our current purposes is that interpretational neutrality requires us to look for criteria for singularity resolution that are applicable whether or not we insist on clock neutrality. That is, the criteria themselves should be defined such that they make sense in a clock neutral or clock non-neutral framework. See \cite[\S5]{gielen:2022} for detailed discussion in a context of a concrete example. That the choices in the quantum representation space and realisations of the basic observables can make a difference for singularity resolution is further evidenced by the analysis of implications of Bohr compactification for singularity resolution studied in \cite{husain:2004}.} We will return to the discussion of this particular issue in Section \ref{sasc}.
 
\section{Observable Based Criteria}
\label{OBC}
\subsection{Bounded Operators}

A linear operator on a Hilbert space, $A: \mathcal{H}\rightarrow\mathcal{H}$, is \textit{bounded} if there is a constant $C$ such that $\|A \psi \| \leq C \|A \| $ for all $\psi \in \mathcal{H}$. By the Hellinger-Toeplitz theorem if an operator is defined on all of $\mathcal{H}$, i.e. has domain $D(A)=\mathcal{H}$, and is symmetric, i.e. such that $<\phi, A \psi>=< A\phi,\psi>$ for all $\psi,\phi \in \mathcal{H}$, then the operator is necessarily bounded. Bounded operators have the attractive property that that exponential of the operator $e^{A}$ is guaranteed to be well defined as a convergent power series.

This means that a direct consequence of the definition of a bounded operator is that such operators necessarily have a unique state independent maximum expectation value. Furthermore, recall that in the Heisenberg picture observables are operator valued functions that take the form of strongly continuous maps from the real numbers in to the space of densely defined, linear, self-adjoint operators on a Hilbert space, $A(t): \mathbb{R}\rightarrow \mathcal{L}_\text{sa}(\mathcal{H})$. Let us then assume that there exists some time independent Hamiltonian operator, $H$, with domain $D(H)$. We then have that if $A(t)$ is bounded and preserves the domain of the Hamiltonian, i.e. $A(t)D(H) \subset D(H)$ for $t\in\mathbb{R}$, then  $A(\cdot)\psi$ is differentiable for any $\psi \in D(H)$ and
\begin{equation}
\frac{d}{dt} A(t) \psi = i [H, A(t)]\psi
\end{equation}
\cite[Proposition 9.2.1]{blank:2008}.
This result means that norm of the operator representing an observable is preserved during time evolution and thus a bounded operator in the relevant class always remains bounded. 

For time dependent Hamiltonians a general condition on the boundedness of the Hamiltonian can be expressed in terms of the restriction that the relevant operator valued function is a strongly continuous map from the real numbers into bounded self-adjoint operators on a Hilbert space, $H:\mathbb{R}\rightarrow \mathcal{B}(\mathcal{H})$. The Dyson expansion:
\begin{equation}
U(t,s)\varphi = 1 + \sum_{n=1}^\infty (-i)^n \int_s^t \int_s^{t_{1}}. . . \int_s^{t_{n-1}} H(t_1) ...H(t_n) \varphi dt_n ... dt_1
\end{equation}
 then allows us to prove the existence of a \textit{unitary propagator} $U(t,s), s,t \in \mathbb{R}$ which is such that:
\begin{eqnarray}
\varphi_s(t) &=&U(t,s) \psi \\
\frac{d}{dt}\varphi_s(t)&=&-iH(t)\varphi_s(t)
 \end{eqnarray}
 where $\varphi_s(s)=\psi\in\mathcal{H}$ and $U(t,s)$ is such that: a) $U(r,s)U(s,t)=U(r,t)$; b) $U(t,t)=I$; and c) $U(s,t)$ is jointly strongly continuous in $s$ and $t$ \cite[Theorem X.69]{reed:1975}. These results provide us with a formal basis under which quantum theories will necessarily have well-posed equations of motion provided the Hamiltonian is bounded in the relevant sense. 

The question is then how best to understand the connection between  boundedness of operators and singularity resolution in quantum cosmology. In particular, for cases in which the Hamiltonian is \textit{not} bounded but other physically relevant operators are. Most directly, we can consider the influential suggestion that non-singular quantum behaviour is indicated by the boundedness of the inverse scale factor \cite{bojowald:2001,bojowald:2001a,bojowald:2002} or inverse 3-volume operator \cite[p.296]{Rovelli:2004}. Such a criterion is prima facie well motivated as a condition for \textit{local curvature} singularity resolution on the basis of the classical relation between curvature divergence and the existence of the inverse metric components. In particular, we might  interpret the finiteness of the eigenvalue of the inverse scale factor operator on states which are annihilated by the volume operator as a quantum resolution of big bang singularity and thus the boundedness of this operator appears as a plausible sufficient condition for singularity resolution \cite{bojowald:2002}. 

In this context it is important to note that the boundedness of kinematical operators, such as the inverse of the 3-volume and inverse scale factor, defined as they are on the kinematical Hilbert space, on which the quantum constraints have not been applied, is not ultimately well suited to act as a sufficient condition to establish non-singular behaviour in the local curvature sense. \textit{Structures defined on the kinematical Hilbert space are not a reliable guide to dynamical possibilities}. Singularity resolution criteria need to be established at the level of physical observables and physical states as defined on the physical Hilbert space to have dynamical significance.\footnote{See for example the discussions of \cite{husain:2004,brunnemann:2006,bojowald:2006,bojowald:2007,bojowald:2011c,ashtekar:2011}.} The point at issue can be made most clear within the internal time framework. In that context, the inverse scale factor is understood to not be gauge invariant and therefore not to be an observable. We can construct a gauge invariant observable based on the 
 inverse scale factor by application of the partial and complete observables technique. However, the spectrum of an operator on the physical Hilbert space can differ drastically from the kinematical spectrum \cite{brunnemann:2006,dittrich:2009}. Thus, kinematic conditions on boundedness of operators are can only ever operate as preliminary guides for potential non-singular behaviour and not as criteria. 
 
The deployment of the property of boundedness of an operator in the context of singularity resolution should then be motivated by the connection between the boundedness of self-adjoint operators on the \textit{physical} Hilbert space and the existence of (state independent) maximum eigenvalues. In particular, for any classical quantity that diverges at the classical singularity we can investigate whether there is a representation of the quantum analogue of this quantity in terms of a bounded self-adjoint operator on the physical Hilbert space. If such a representation exists then we are guaranteed that the expectation value converges to some finite value in a state independent manner. 

Following \citeN{ashtekar:2011}, and again expressing things in terms of the internal time view, the idea is to construct a physical Hilbert space and a complete family of Dirac observables at least some of which diverge at the singularity in the classical theory.\footnote{For early work on dynamical singularity resolution in Loop Quantum Cosmology see \cite{ashtekar:2006a,ashtekar:2006b,ashtekar:2006c,ashtekar:2007b,ashtekar:2008,singh:2009}.} For example, the matter density, anisotropic shears and curvature invariants evaluated at an instant of a suitably chosen relational time. We can then ask: do the corresponding operators all remain bounded on the physical Hilbert? If the answer is yes, then the singularity is resolved locally in a physically reliable sense.  This provides us with a highly plausible sufficient condition for local curvature singularity resolution:
\begin{itemize}
\item [\textbf{SR1}:] \textit{Bounded and Self-Adjoint Physical Operators}. Given a classical quantity that diverges at the singularity, the existence of a quantum representation of this quantity as a bounded and self-adjoint operator on the physical Hilbert space is a \textit{sufficient} condition for \textit{local curvature} singularity resolution.   
\end{itemize}
An example of the successful application of this criterion in the context of an explicit model is provided by treatment in exactly solvable isotropic Loop Quantum Cosmology model with massless scalar as the matter content \cite{ashtekar:2008,singh:2009}. In particular, it is shown that on the physical Hilbert space the energy density operator is bounded with a critical value $\rho_\text{crit}$ the supremum of the spectrum.\footnote{There are a range of similar results indicating the robustness of resolution in the sense of SR1 in Loop Quantum Cosmology. For further details and references see \cite{ashtekar:2011} and \cite{ashtekar:2021}. For a more sceptical critical commentary see \cite{bojowald:2020}.}  

Such examples give a degree of prime facie physical plausibility to SR1 as a sufficient condition for local curvature singularity resolution. What of the global dynamics notion? Most obviously, in order to extend the condition towards a global dynamics notion of singularity resolution it seems plausible to consider whether additional conditions could be placed regarding the Hamiltonian. In particular, we might consider whether we could construct a condition for global dynamics singularity resolution by requiring that the Hamiltonian is also bounded and thus a unitary propagator is available directly via the Dyson expansion. However, in cosmology one expects physical Hamiltonians to be unbounded or semi-bounded, and thus such a strengthened version of SR1 would be highly limiting with regards to physical applicability. SR1 is too stringent to be useful as a sufficient condition for global dynamics singularity resolution in cosmology. In a physical context, what we would like to consider is whether there is a weaker condition based upon the requirement for Hamiltonians to at least be self-adjoint, if not bounded. We defer discussion of this important aspect of singularity resolution to our treatment of unitarity and inner product based criteria in \S \ref{unitarity}.

The final question worth considering in the context of bounded operators and criteria for singularity resolution is whether or not boundedness of operators on the physical Hilbert space might plausibly also be asserted as a necessary condition for local curvature singularity resolution. That is, we could then postulate that given a classical quantity that diverges at the singularity, the existence of a quantum representation of this quantity as a bounded and self-adjoint operator on the physical Hilbert space is a \textit{necessary and sufficient} condition for singularity resolution. 

Promulgation of such a stronger criterion is under-justified however. This is for several reasons. First, often we are interested in the quantum behaviour of an operator that is classically zero, and thus the salient physical question for singularity resolution is whether or not the expectation value vanishes rather that whether it has a finite limit and is thus bounded. Thus the existence of a non-zero but state dependent minimum could be taken to be enough to justify a claim of singularity resolution and this can of course obtain for unbounded operators. Second, as already noted, unbounded operators are ubiquitous in quantum theories. It is thus natural to look for a criterion for singularity resolution that is applicable to such objects as well. Third and finally, ultimately there being a \textit{state-independent} finite convergence of the expectation value in the quantum theory as the classical big bang singularity is approached does not have a compelling physical motivation. It is thus sensible to consider weaker candidates for possible necessary and sufficient conditions for singularity resolution that are tied to the behaviour of the expectation values irrespective of whether or not salient operators are bounded.

\subsection{Non-Singular Expectation Values}

Analysis of the behaviour of expectation values is a standard means of tracking physically salient properties of a quantum theory. In particular, the semi-classical analysis of a quantum theory typically relies on the connection between the behaviour of the expectation values and the classical equations of motion. The idea that we should focus on expectation values in the analysis of singularity resolution is thus naturally aligned with the idea of classical correspondence. The most obvious criterion that can be constructed based upon this connection is a quantum analogue of the classical local curvature singularity criterion of curvature blow up understood in terms of the direct correspondence between curvature invariants growing without bound in finite time and the formal divergence of relevant expectation values. This suggests the following necessary criterion:
\begin{itemize}
\item [\textbf{SR2}:] \textit{Finite Expectation Values}. Given a classical theory in which some class of classical observable diverges it is \textit{necessary} for \textit{local curvature} singularity resolution that the expectation values of the corresponding quantum operators are always finite, if they are ever finite. 
\end{itemize}
This criterion is explicitly defended as a necessary condition in \cite{Gryb:2017a} and occurs in slightly different forms in various discussions in the context of the requirement for finiteness of expectations of energy, e.g., density \cite{brunnemann:2006}.

A straightforward attractive feature is that the approach is evidently interpretational neutral with regard to the problem of time at least. That is, the finite expectation value criterion is applicable both to Schr\"{o}dinger-type models, where the expectation values are considered at all values of the extrinsic time parameter, and to Wheeler--DeWitt-type models, where the expectation values are projected onto a physical Hilbert space and are considered at all values of some internal time.\footnote{There is, however, a subtlety here relating to range of internal time parameters. In particular, within Wheeler--DeWitt-type models an infinite range of an
internal time might correspond to a finite range of proper time. This means that even if expectation values stay finite for all values of an internal time, they may asymptotically diverge. We will return to this issue in the context of the discussion of slow vs fast clocks in \S \ref{sasc}.} 

One might worry, however, that a focus on expectation values will lead to a difficulty with SR2 and interpretational neutrality in the context of the measurement problem. In particular, following \cite{tipler:1986}, one might argue that any criterion based on  expectation values relies upon the possibility of repeated measurement. This would seem to imply that we require an ensemble of universes to give physical meaning to the expectation value. The use of expectation values in quantum cosmology is then, at the least, inconsistent with any interpretation in which only one universe available, and possibly simply interpretationally inconsistent simpliciter. Tipler's argument  not only seeks to connect use of expectation values to an operationalist interpretation but to demonstrate that such an interpretation can be straightforwardly shown to quickly degenerate into physical absurdity in the context of quantum cosmology. 

The simple response to this worry is that expectation values are \textit{not} defined to be the average of a repeated measurements. Rather, such an understanding itself comes from a particular interpretation of the quantum formalism in a particular context. There is no necessity of applying an interpretation of expectation values as averages of repeated measurements in the context of quantum cosmology. Rather, from an interpretationally neutral standpoint we can and should focus on the more abstract and basic status of expectation values within the quantum moment expansion which can be used to define effective equations of motion.\footnote{A further fascinating line of response due to Bojowald (personal communication) would be to appeal to the BKL conjecture to recover the possibility of a statistical interpretation of expectation values. In particular, we could consider a single inhomogeneous universe, modelled by a homogeneous minisuperspace model. We then conceive of the model via an ensemble of
approximately homogeneous patches. Then, invoking the BKL conjecture, these patches can be expected to evolve largely independently of one another. We thus arrive at precisely the ensemble of universes needed to make sense of a statistical interpretation of expectation values. On a slightly different note, for work on the general problem of understanding probability in quantum cosmology in the context of the consistent histories approach see \cite{craig:2010,craig:2013}.}

For example, using the notation of \cite{bojowald:2006b}, the generalized moments for a 4-dimensional classical phase space, $(q^1, q^2, p^1, p^2)$, can be expressed as
\begin{multline}\label{eq:moments defn}
  G^{a_{k1}, a_{k2}}_{b_{k1},b_{k2}} = \mean{ \lf( \hat q^{k_{1}} - q^{k_{1}}\rt)^{a_{k1}}\lf( \hat q^{k_{2}} - q^{k_{2}}\rt)^{a_{k2}} \rt. \\ \lf. \times \lf( \hat p^{k_{1}} - p^{k_{1}}\rt)^{b_{k1}}\lf( \hat p^{k_{2}} - p^{k_{2}}\rt)^{b_{k2}} }_\text{Weyl}\,,
\end{multline}
for $ {{a_{ki}}, {b_{ki}} } = 0, 1, \hdots, \infty$. The Weyl subscript indicates completely symmetric ordering. The evolution equations and commutation relations for the system can be expressed in general terms as symplectic flow equations for these moments.\footnote{For the explicit statement of these equations, see \cite{bojowald:2006b}.} 

The finiteness of expectation values is then understood as part of a more general requirement for the existence of well defined effective equations of motion as given by quantum corrections to Hamilton's equations for the expectation values \cite{bojowald:2006b,brizuela:2014}. These corrections can be explicitly characterised in terms of the quantum moments associated with the system.\footnote{For analysis of Wheeler-DeWitt type quantum cosmology and Loop Quantum cosmology within the framework of effective theory based upon moment expansion see \cite{bojowald:2007b,bojowald:2011b,bojowald2012c}. For discussion of related issues in the context of Group Field Theory see \cite{marchetti:2021b}.}  Thus, appealing to finiteness of expectation values as a necessary condition for singularity resolution need not violate interpretational neutrality, rather the criterion can be understood as the first step towards the requirement the full set of quantum corrections to the classical theory will be finite for all times. We could thus strengthen the necessary condition to get a condition SR2$^\star$:
\begin{itemize}
\item [\textbf{SR2$^\star$}:] \textit{Finite Moment Expansion}. Given a classical theory in which some class of classical observable diverges it is \textit{necessary} for \textit{local curvature} singularity resolution that the quantum moment expansion is always finite, if it is ever finite.
\end{itemize} 
This is a stronger necessary condition for local curvature singularity resolution since we have that SR2$^\star$$\rightarrow$SR2 but SR2$\nrightarrow$SR2$^\star$ and might be preferred on that basis. We will not take a position of this here and will often leave implicit in what follows that expectation values or moment expansions might be considered interchangeably.\footnote{It is worth noting that there is a plausibly worry that SR2$^\star$ may be too strong since divergence in the moment  expansion may also be associated with a state no long being in the domain in which an unbounded operator used to define the moments can be applied. Arguably such a divergence would be due to issues with representation rather than a physical quantum correlate of the classical singularity. Such issues suggest caution in applying SR2$^\star$ as a necessary condition.} The idea of well defined effective equations of motion will be further discussed in \S \ref{QH} in the context of the idea of \textit{quantum hyperbolicity} as a candidate condition for local curvature and global dynamics singularity resolution. 

Let us now consider whether finiteness of expectation values -- or more generally the moment expansion -- might also form the plausible basis for a sufficient condition for local curvature singularity resolution. In other words, can a quantum theory in which the quantum moments are finite to all orders ever be considered singular? Classical correspondence offers us a mixed verdict here.  On the one hand, finiteness in the relevant sense would break the direct correspondence to curvature blow up and thus plausibly be sufficient for local curvature singularity resolution, in the sense that we do not have an analogue of curvature invariants growing without bound in finite time. On the other hand, we might consider there to be two importantly different behaviours, both of which can obtain in a finite quantum theory; that is, quantum evolution that follows the classical singular behaviour `to infinity' whilst remaining finite, and quantum evolution that diverges from the singular classical behaviour in an explicit sense. 

We might then focus on the relative difference between expectation value and classical value as the salient quantity. For example, for matter density we could consider the quantity:

\begin{equation}
\triangle_\text{rel} (\rho)= \big{(} 1 - \frac{ <\rho > }{\rho }\big{)} 
\end{equation}  

Consider then a quantum theory in which the moment expansion is finite to all orders and we thus get well defined evolution of all physical quantities at all times. If we were to find that $\triangle_\text{rel}$ remained small as the classical value grows without bound in finite proper time along an incomplete inextendible causal curve, then we might plausibly still consider the singularity unresolved, notwithstanding the fact that the quantum evolution equations never break down and there is formally no divergence. In fact, as discussed further in \S \ref{unitarity}, recent work by \citeN{gielen:2022} has shown that it is possible to construct unitary quantum cosmological models with precisely such properties. If we would like to classify such models as displaying (local curvature) singular behaviour then we must look for a stronger sufficient condition for singularity resolution than finiteness of expectation values or the moment expansion. Thus SR2 and SR2$^\star$ are best understood as necessary conditions only. 

The most natural option for a sufficient condition, widely discussed in the literature, is to focus on vanishing of expectation values  for quantum representations of classical quantities that vanish at the big bang. Whereas well defined quantum evolution equations will `protect' expectation values from formally diverging, even in an infinite time limit, it is possible to for expectation values to converge to zero. The key observable here is the volume operator. This can be considered explicitly or via the behaviour of the scale factor or some other suitably connected geometric variable such as the radius of the universe squared. 

The central idea is that we have a variable which is such that its classical value zero is directly connected to the big bang singularity. Zero volume generically plays that role in FLRW cosmologies including mini-superspace models. The physical connection to the local curvature singularity and curvature blow-up is based on the assumption that the variable representing stress-energy within the model will be an integral of motion (or at least non-vanishing) and thus we have singular behaviour as $v\rightarrow 0$ since we can then understand the semi-classical energy density given by a ratio of expectation values as divergent. 

This leads to a the following putative sufficient condition for singularity resolution:
\begin{itemize}
\item [\textbf{SR3}:] \textit{Non-Zero Expectation Values}. Given a classical theory in which the singular behaviour is associated with some set of classical observable vanishing, it is \textit{sufficient} for \textit{local curvature} singularity resolution that the expectation value of the corresponding quantum operators are non-zero for all quantum states. 
\end{itemize}
This criterion was originally proposed in the discussions of \cite{lund:1973,gotay:1980,gotay:1983} and is considered within various modern discussions -- see in particular \cite{gielen:2020,gielen:2022}.\footnote{A similar idea is also discussed by \cite{ashtekar:2008}. That treatment does not assert non-zero expectation values as a necessary or sufficient condition but rather asserts that it is a `rather weak criterion' that is, however, applicable to any state and thus a good means of `testing' theories for non-singular behaviour. This treatment is thus very much along the lines of an indicative condition and is one we shall ultimately endorse.} 

Concrete demonstrations of behaviour that displays a local curvature singularity in the sense of vanishing expectation values generally focuses on the volume operator and reiles on a choice of internal time which is such that classically $v\rightarrow 0$ as $t\rightarrow\infty$. For example, in \cite[\S IV.A.]{ashtekar:2008} an internal clock is chosen based upon a scalar field with a mini-superspace model. The Dirac observable $V_\phi$ can then be constructed and its classical singular behaviour identified with the  $\phi \rightarrow \pm \infty$ limit.\footnote{The relevance of the idea of requiring an infinite internal time limit in the context of a unitary evolution will be discussed in \S\ref{sasc}.} Crucially it can then be shown that $<V_\phi>\rightarrow 0$ as $\phi \rightarrow \pm \infty$ (where within which of $\pm$ depends on which branch of solution is chosen). By contrast, resolution would occur in the sense of SR3 if we were to find that $<V_\phi>$ attains a  non-zero minimum precisely as $V_\phi$ goes to zero in the classical theory. 

A number of attractive qualities of this understanding of local curvature singularity resolution are worth mentioning. First, in general the minimum value of the volume operator that is indicative of non-singular behaviour need not be state independent since we have not restricted to bounded operators. The criterion is strong enough however to exclude resolution being state dependent in that it does not lead us to classify as non-singular models which have non-zero volume expectation values only for special initial states. Second, there is a clear correspondence between this sense of local curvature singularity resolution and the idea of relative difference between classical and quantum evolutions being significant. In particular, satisfaction of SR3 implies the quantity $\triangle_\text{rel} (V)$ always remains small for singular models and may grown without bound for non-singular models (in particular for those that display `bouncing' behaviour). Third, and most significantly, there is a plausible connection between our ability to formally demonstrate the vanishing of the expectation value and the choice of internal time in which the singularity is associated with an infinite value of the variable that plays the role of the clock. This feature has lead to the conjecture that such `fast clocks' are always singular when unitary. We will return to this idea in detail in \S\ref{sasc} and where we return to the connection between the fast clock conjecture and SR3 in the specific context of unitary quantum dynamics. 

Notwithstanding these attractive features, there is good model based evidence \textit{not} to treat SR3 as a constitutive condition for local curvature singularity avoidance. Consider in particular pressure singularities in anisotropic Bianchi models in which curvature invariants diverge at finite proper time and non-zero volume. This feature can be seen explicitly for the case of the Loop Quantum Cosmology treatment of Bianchi I models with vanishing anisotropic stress due to \citeN{singh:2012}. In particular, it is shown that the Ricci and Weyl scalars can diverge at non-zero volume and finite energy density for certain choices of matter field on account of a pressure divergence. For this reason, SR3 can not be understood as a sufficient condition for local curvature singularity avoidance in any straightforward sense. 

There is a subtlety, however, regarding whether what is issue is whether we are focused on big bang singularity avoidance or local curvature singularity avoidance simpliciter. As noted earlier, the big bang singularity is an example of a strong singularity in which integrals of curvature invariants diverge as a critical value of the affine parameter is approached and any physical detector is inevitably destroyed. The model based evidence from the study of a range of Bianchi models in the loop quantum cosmology framework indicates that in these models at least we \textit{do not} find strong local curvature singularities at non-zero volume expectation value \cite{singh:2012,saini:2017,saini:2018}. Interestingly, it is further shown in these studies that whereas in the classical treatment of the Bianchi models in question one finds a strong singularity associated with geodesic incompleteness and thus a global dynamics singularity, in the quantum model there is an absence of a strong singularity together with well defined effective dynamical equations. This is in line with the quantum extension of the classical conjecture that strong singularities might be necessary and sufficient for geodesic incompleteness. \cite{krolak:1987,tipler:1980}. It seems plausible therefore to consider SR3 as in fact being (at least) an \textit{indicative} condition for \textit{global dynamics} singularity avoidance. We will return to the connection between vanishing expectation values and indicative conditions for global dynamics singularity resolution in \S\ref{sasc}.      

To what extent are the conditions SR1-3 mutually supporting? The internal relation between SR1 and SR2 is fairly direct since by definition all bounded operators must have finite expectation values, operators with finite expectation values need not be bounded. We thus have that SR2 is necessary but not sufficient for SR1. Intuitively we would expect a similar relation to hold between SR3 and SR1 or SR2 in that one might expect the vanishing of the expectation value of the volume operator to be associated with the divergence of the matter density and, conversely, the boundedness of the matter density to be associated with non-zero expectation value of the volume. These connections are, however, not so straightforward and there is evidence for the possibility of models in which the matter density is bounded whilst the zero volume expectation value is reached \cite{bojowald:2020}. Such possibilities reflect the fact that the zero of expectation value of volume is an aggregate and so does not imply any particular matter density operator to involve a finite energy within a zero volume. 

Moreover, the physical motivation for adopting vanishing volume expectation value as a sufficient condition for singularity resolution is ultimately based on the assumption that singular models are such that the expectation value of the volume cannot evolve beyond the zero value and there is thus a tracking between classical and quantum singular dynamics rather than bouncing phenomenology. However, there is nothing that forbids a quantum bounce being combined with zero volume expectation value. The plausibility of such a possibility is made particularly clear by considering the possibility of superpositions between histories in which the volume operator does and does not go to zero. Nothing seems to forbid the expectation values or moments being finite in such cases and thus SR2 or SR2$^\star$ may hold when SR3 fails. Ultimately, such possibilities point to a further reason that we should think of SR3 as an indicative rather than constitutive condition.  

\section{State Based Criteria}

\subsection{Vanishing Wavefunction}
\label{vanishingwf}

We now move on to consider criteria for \textit{global dynamics} and \textit{local curvature} singularity resolution based upon the behaviour of the quantum state. Our first candidate criterion is the longstanding and intuitive condition based on the idea of the wavefunction vanishing for singular configurations. 

The general idea of appealing to the vanishing of the wavefunction as a condition for a non-singular quantum theory is associated with the early analysis of quantum mini-superspace by \citeN{DeWitt:1967}. In addition to this impressive pedigree, the idea has intuitive motivation based on the connection between the wavefunction and probability. Furthermore, there has been a recent revival of interests in the criterion in the context of a conformally invariant reformulation \shortcite{Kiefer:2010,Kiefer:2019,kiefer:2019b}. We will return to this treatment shortly. In the meantime, let us explicitly formulate the criterion as follows:

\begin{itemize}
\item [\textbf{SR4}:] \textit{Wavefunction Vanishing}. It is a  \textit{sufficient} condition for \textit{global dynamics} singularity resolution that the wavefunction has zero support on three-geometries of zero volume.
\end{itemize}
That wavefunction vanishing might only plausibly be taken as a sufficient and not necessary criterion for global dynamics singularity resolution is straightforwardly demonstrated by the example of quantum systems which avoid classical singularities despite the wavefunction \textit{not} vanishing. For example, the ground state solution of the Dirac equation for the hydrogen atom \cite{kiefer:2019b}.

Before we proceed with our analysis it is worth noting from a historical perspective that it does not seem entirely correct to attribute to DeWitt a full throated assertion of the candidate criterion. He introduces the idea as follows:
\begin{quote}
[T]he quantum physicist may be able to dispose of [the singularity] by simply imposing, on the state functional, the following condition:
\begin{equation}
\Psi [^{(3)}\mathcal{G}]=0 \textrm{ for all } ^{(3)}\mathcal{G} \textrm{ on } \mathcal{B}_Q
\end{equation}  
\textit{Provided it does not turn out to be ultimately inconsistent, this condition}, [...] yields two important results. Firstly, it makes the probability amplitude for the catastrophic 3-geometries vanish and hence gets the physicist out of his classical collapse predicament. Secondly, it may permit the Cauchy problem for the [Wheeler-DeWitt equation] to be handled in a manner very similar to the Schr\"odinger equation [in that] specification of $\Psi$ on $\Sigma$ together with the boundary condition [above] suffices to determine $\Psi [^{(3)}\mathcal{G}]$ completely for all $^{(3)}\mathcal{G}$.    
\end{quote}
Here $^{(3)}\mathcal{G}$ is a three geometry (in superspace) and $ \mathcal{B}_Q$ represents the `quantum barrier' constituted by a zero volume 3-space. 

Two important qualifications are thus made within DeWitt's analysis. First, he admits the possibility that the condition may simply be inconsistent and thus yield no content. Second, he suggests that the condition \textit{may} allow for well defined quantum evolution beyond the quantum barrier along the lines of a global dynamics condition. Thus, DeWitt's analysis makes clear that the merely vanishing of the probability amplitude on the classically problematic 3-geometries is not sufficient for quantum global dynamics singularity resolution per se -- rather what is needed is well defined quantum evolution equations with a solvable Cauchy problem. We can plausibly understand DeWitt here as indicating the vanishing wavefunction boundary condition as an indicative condition for global dynamics resolution, but in fact advocating for well-defined equations as the relevant constitutive condition. Such an interpretation would then imply that the `DeWitt criterion' should really be something closer to that advocated by Bojowald in the context of his quantum hyperbolicity condition which will be discussed in \S\ref{QH}. 

Let us now consider the plausibility of the criterion of wavefunction vanishing as a sufficient local curvature condition with the attribution of the idea to DeWitt suitably bracketed. An influential analysis of the idea can be found in an early study of quantized mini-superspace due to \citeN{blyth:1975}. It is argued that the absence of a local curvature singularity in the model in question, and more generally, cannot be guaranteed by the vanishing of the wavefunction. In particular, the fact that $\psi(a)=0$ for the scale factor $a=0$, is not understood as sufficient for singularity resolution since if $a$ has a purely continuous spectrum, as it does in the model considered, then it is also taken to be required that the spectrum contains an isolated point which corresponded to the singular value. Furthermore, what is important, according to \citeN[p.774]{blyth:1975}, is the behaviour of the relevant probability density as the classically singular point is approached. In particular, what we should consider is the transition rate from the initial state into the various `singular' states and not merely the long term behavior of the state function. Relatedly, in a more recent criticism of SR4, \citeN{ashtekar:2011} note that vanishing of the wavefunction at $a=0$ does not guarantee that the physical inner product would be local in $a$ nor exclude the possibility that matter density and curvature may still grow unboundedly at early times. 

The key point is that ultimately the study of quantum correspondence to classically dynamical behaviour must feed into a physically reliable local curvature or global dynamics criterion of singularity resolution.  As recognised by DeWitt, the behaviour of the wavefunction alone is neither necessary nor sufficient to \textit{guarantee} non-singular dynamical behaviour in either the local curvature or global dynamics sense. Thus vanishing of the wavefunction is best understood as an indicative condition, rather than a constitutive criterion for singularity resolution.

A modern reformulation of SR4 has been provided in the treatment of \citeN{kiefer:2019b}. The motivation for this approach is the observation that the SR4 appears to break the manifest invariance of the theory. In particular, the criterion fails to be conformally invariant in three or more dimensions. This leads to the suggestion that the criterion be adjusted to be manifestly conformally invariant through expression as a density of conformal weight zero. This leads to the following criterion:
\begin{itemize}
\item [\textbf{SR5}:] \textit{Conformally Invariant Wavefunction Vanishing}. It is a  \textit{sufficient} condition for global dynamics singularity resolution that the expression $\star |\Psi|  ^{\frac{2d}{d-2}}\rightarrow 0$ in the vicinity of the singularity. 
\end{itemize}
where $\star$ is the Hodge star and $d$ is the dimensionality of the model. 

It remains to be seen whether the conformally invariant wavefunction vanishing criterion fairs any better than the original proposal. One issue is that the notion of `vicinity' in SR5 needs to be made precise via reference to a suitable limit and topology. A further issue, discussed at length by \citeN{Warrier:2022}, is with regard to stability of applicability of the criterion. In particular, the inclusion of the dimensionality within the reformulated criterion itself appears to render it unstable under extension to infinite dimensions. This suggest, at the least, that what is required is an even more general formulation of the criterion based upon a generalisable expression for the vanishing of density of conformal weight zero that is well defined in the field theoretic case.   

\subsection{Quantum Hyperbolicity}
\label{QH}
Following \citeN[pp.305-6]{bojowald:2007}, the idea of quantum hyperbolicity is to generalise the classical conditions for well-posedness of the partial differential equations describing the dynamics of matter to the fundamental object of quantum space-time itself. The core idea is that the quantum state should be `extendible' across or around all sub-manifolds of classically singular configurations. By contrast, in a situation in which a quantum state defined on the space of metrics cannot be `extended' uniquely across a given classically singular sub-manifold, Bojowald argues we should understand there to be a boundary to the quantum evolution and we should take there to be a certain `incompleteness' of the quantum space-time. 

There is strong degree of natural connection between quantum hyperbolicity as Bojowald defines it and classical singularity conditions. This connection can, in particular, be drawn in terms of both well-posedness of the equations of motion and incompleteness of geodesics and thus the global dynamics sense of singularity. We should note, however, that the connection between the classical notions of global hyperbolicity and spacetime (in)extenablity is a complex one and the terminology of `extending' the quantum state in the context of a definition of quantum hyperbolicity might thus be misleading. For this reason we will slightly reformulate Bojowald's original definition, whilst keeping the same spirit, by focusing on the well-posed quantum evolution equations. With this in mind, let us consider the formulation of quantum hyperbolicity as a necessary and sufficient criterion for quantum singularity resolution as follows:
\begin{itemize}
\item [\textbf{SR6}:] \textit{Quantum Hyperbolicity}. It is a  \textit{necessary} and \textit{sufficient} condition for \textit{global dynamics} singularity resolution that the quantum state should have unique and well-posed evolution equations across or around all sub-manifolds of classically singular configurations.
\end{itemize}
This formulation is extremely general. In practice, the assessment of the result of SR6 would require a detailed analysis in terms of the following steps \cite{bojowald:2007}: First, the diagnosis of the phase space locations of the classical singularities need to be determined. Second, the relevant division of the classical space of metrics into disconnected components with and without singular structure has to be represented. Third, a means of representing the evolution of the quantum state across the singular region needs to be provided. This last step would require a clear connection to be drawn between the phase space representation of the singularities and the domain of the quantum state. It is only after completion of these three steps that we can make an assessment of resolution or not be made in terms of the relevant dynamical equations describing the extension and whether or not such an evolution obtains generically or only for special initial conditions. 

Despite its manifest formal and physical cogency the criterion of quantum hyperbolicity (SR6) has not been widely applied in treatments of singularity resolution subsequent to the \citeN{bojowald:2007} formulation. This is not perhaps surprising due to the high computational demands required in its assessment. This is particularly true in the context of the specific formulation the third step of evolution via internal time approaches. In particular, on internal time approaches evolution requires de-paramterization in terms of phase space variable that is monotonic around the classical singularity. Formulation of the evolution equations is thus both calculationally intensive and representation dependent in a fundamental sense. Although undoubtedly challenging, detailed consideration of the specific hyperbolic quantum dynamics of particular models or family of is likely to be of significant value it terms of establishing connections to local curvature criteria, such as SR1. Such a project will not be attempted here however. Rather, we will consider the more general question of the relation between clock dependence and singularity resolution in \S\ref{sasc}. 

What is important to note in the context of current discussion is that outside the context of internal time approaches it is not clear that the full demands of quantum hyperbolicity will be required. As discussed in \cite{Gryb:2017a}, in an internal time context we typically have a Wheeler--DeWitt-type equation where the physical Hilbert space on which evolution takes place is represented in terms of operators acting on functions whose domains are ever changing submanifolds, parametrized by internal time, of the configuration space. Here, the dynamics can force the system directly through a classically singular submanifold, and this can lead to potential problems that quantum hyperbolicity is well-suited to diagnose. In contrast, in the case of an extrinsic time model, the domain of the relevant functions remains fixed to the entirety of the original configuration space. Because the classically singular region is a set of measure zero on this space, the well-posedness of the equations of motion can be anchored simply in considerations of boundedness and unitarity. There is thus a plausible formal and pragmatic basis to consider conditions based upon unitarity alone rather than focusing on the full demands of quantum hyperbolicity. Such inner-product based criteria are the focus of the following, final section. 

\section{Inner Product Based Criteria}
\label{unitarity}

\subsection{Essential Self-Adjointness and Unitarity}
\label{unitarity}

Unitarity is a fundamental assumption made by researchers in most areas of quantum physics and quantum cosmology is no exception. The connection between unitarity, self-adjointness, and singularity resolution is a subtle one, however. By definition a unitary operator is a bounded linear operator on a Hilbert space that is surjective and preserves the inner product. For our purposes the most physically insightful approach to the diagnosis of the unitarity of a quantum cosmological model is to focus on the conservation of the inner product under the transformation generated by an intrinsic or extrinsic time evolution: a surjective evolution map associated with a bounded linear operator is then defined to be unitary if and only if it preserves the inner product.  

For systems with a time independent Hamiltonian, unitarity in this sense can be expressed as follows. Consider a surjective evolution map associated with the bounded linear operator $U_t: |\Psi (0)> \rightarrow |\Psi (t)>$ for $\Psi \in |H>$. Unitary then obtains if and only if: 
\begin{equation}\label{equnitarity} 
\frac{d}{dt} <\Psi | \Phi >_t = 0 
\end{equation}
where the subscript $t$ indicates that the inner-product may depend upon which variable is chosen as the clock in intrinsic time approaches. For theories in which we can construct the dynamics via a Schr\"odinger-type equation the evolution map is then required to solve the equation:  
\begin{equation}\label{eveq}
H_t |\Psi> = i \hbar  \frac{d}{dt} |\Psi >
\end{equation}
A candidate unitary evolution map that uniquely solves (\ref{eveq}) is then automatically given by the exponentiation of the Hamiltonian in the form $U(t) = e^{-iH_t t}$. Crucially the evolution operator given by exponentiation of a Hamiltonian may be bounded even when the Hamiltonian operator is not. 

Unitarity thus depends on the properties of the physical Hamiltonian operator understood to generate intrinsic or extrinsic time evolution. In intrinsic time approaches this will \textit{not}, of course, be the full Hamiltonian given by the relevant Wheeler-DeWitt operator but rather a de-parameterization thereof. We will consider the crucial connection between the existence and uniqueness of a unitarity evolution map and self-adjointness of the physical Hamiltonian shortly.

The extension of our analysis of unitarity to cases where the classical Hamiltonian is time-dependent, and thus we have an operator valued function rather than a single Hamiltonian, is in general non-trivial. In the case of bounded Hamiltonians this extension is straightforward since the Dyson expansion then allows us to prove the existence of a unitary propagator which gives a unique solution the relevant time-dependent Schr\"odinger equation. In contrast, in the physically more important case of unbounded Hamiltonians, there does not exist such strong general results and proofs of the existence of a unitary propagator and well-posed equations of motion typically require  assumptions regarding the specific form of the Hamiltonian \cite[\S9.5]{blank:2008}. 

Here we will neglect the details of such a full analysis of the existence and uniqueness of solutions to the time-dependent Schr\"odinger equation for unbounded Hamiltonians and its connection to quantum cosmological models. Rather, it will prove sufficient to rely on the time-independent analysis given the assumption that the propagator in question can be constructed via products of operators defined at each time step. We would then have a family of Hamiltonians and evolution operators one for each time step and be able to understand the unitarity of each operator in terms of the conservation of the inner product at that time step. See \cite[\S V]{gotay:1983} for an explicit construction along these lines.

It is natural then to consider the connection between unitarity and the self-adjointness of the (time independent) operator  $H_t $. To do this we must first distinguish between three different formal possibilities with regard to the self-adjointness of an arbitrary dense, symmetric operator $A$ with domain $\mathcal{D}(A)$. The three possibilities can be formulated in terms of the existence and uniqueness of \textit{extensions} to the domain of the operator, $\mathcal{D}(A) \subseteq  \mathcal{D}(A^\dagger)$. In particular, let us use introduce a terminology in which we define \textit{essentially self-adjoint operators} as those operators with a unique extension, \textit{extendible operators} as those operators with an infinite family of parameterised extensions, and \textit{non-extendible operators} as those operators do not have any self-adjoint extensions.  

A theorem, due to von~Neumann, allows us to diagnose into which of the three categories a particular symmetric operator falls  based upon the dimensionality of the relevant \textit{deficiency subspaces} -- see \citeN{reed:1975}, theorem X.1. A further theorem, also due to von~Neumann, implies that all symmetric operators that commute with complex conjugation are either essentially self-adjoint or extendible -- see \citeN{reed:1975}, theorem X.3. Finally, Stone's theorem for one-parameter unitary groups then implies a one-to-one  correspondence between self-adjoint operators and strongly continuous one–parameter unitary groups  \cite[VIII.4]{reed:1980}. Significantly, the  self-adjoint operators in question may be either essentially self-adjoint or particular members of the family of self-adjoint extensions of some extendible operator. 

The immediate implication of Stone's theorem together with the second von~Neumann theorem is that for physical systems the existence of a unitary evolution map is all but guaranteed. This is because we can generally presume the form of physical Hamiltonians to be such that they are symmetric operators which are purely real in the sense that they do not have any imaginary parts. This means that so long as it is symmetric, we are guaranteed the existence of a (possibly non-unique) self-adjoint extension of the Hamiltonian operator which induces a corresponding unitary evolution map via exponentiation. In finite dimensional quantum cosmological models at least we generally do have symmetric Hamiltonians and thus can expect unitary evolution maps to exist. The issue is then for Hamiltonians that are not-essentially self-adjoint how do we find the extensions and what interpretation should we attached to them. 

It is in this context that the deficiency subspaces framework for the analysis of self-adjointness proves not to be particularly physically insightful. In particular, it is useful to be able to establish a more direct connection between the self-adjoint extensions, physical boundary conditions, and unitarity. Let us then focus our attention on the boundary conditions that need to be imposed on states in the Hilbert space such that the condition \eqref{equnitarity} obtains. This method can be shown to be equivalent to the von~Neumann treatment \cite{gitman:2012}. Focusing our attention on self-adjointness of the Hamiltonian operator, which we assume to essentially self-adjoin or extendible, then leaves two possibilities:
\begin{enumerate}
	\item [i.] All states, which are square integrable with respect to the appropriate interval and measure, automatically satisfy the boundary conditions. In this case, the Hamiltonian operator is \emph{essentially self-adjoint}. 
	\item [ii.] All square integrable functions are spanned by the solutions of a unitary family of boundary conditions. This implies an underdetermination in the self-adjoint representations of the operator in question. The Hamiltonian is an \textit{extendible operator} and the solution space of each member of the unitary family defines the domain of a particular self-adjoint extensions.
\end{enumerate}
Significantly Stone's theorem implies that in both case i. and ii. the dynamics will be unitary. In the first case, unitarity is automatic and the evolution map is identified with a single continuous one-parameter unitary group. In the second case, unitarity depends on the imposition of the boundary conditions and the unitary family of boundary conditions corresponds to a parameterised family of evolution maps each in turn identified as continuous one-parameter unitary groups. 

Two candidate criteria for singularity resolution can then be distinguished based upon the cases i. and ii. The logically strongest approach that can be found in the literature is due to \citeN{horowitz:1995}. The essence of 
Horowitz and Marolf proposal is that, in general, a system is non-singular when the evolution of any state is uniquely defined for all times. By contrast, for a singular system, we will find a `loss of predictability'. The particular connection with essential self-adjointness is then argued for on the basis of a correspondence between the existence of non-unique self-adjoint extensions to the Hamiltonian in the quantum theory and the presence of incomplete geodesics in the classical theory on the particular basis of the joint requirement for a choice of boundary conditions \cite[p. 5670]{horowitz:1995}. This is of course to focus on the global dynamics sense of singularity resolution. We can thus reasonably understand Horowitz and Marolf to be proposing the following criterion:
\begin{itemize}
\item [\textbf{SR7}:] \textit{Essential self-adjointness} of the Hamiltonian is a \textit{necessary and sufficient} condition for \textit{global dynamics} singularity resolution in quantum cosmology.
\end{itemize}
In this definition we should interpret Hamiltonian as referring to the physical Hamiltonian that generates evolution of the quantum state with respect to an intrinsic or extrinsic time. Essential self-adjointness of the Hamiltonian then corresponds to the uniqueness and unitary of the dynamical map generated by the Hamiltonian via Stone's theorem.

How plausible is SR7? The most obvious issue is with regard to classical correspondence. The global dynamics sense of classical singularity is the existence of incomplete \textit{and inextendible} paths. By contrast, the initial example of incomplete geodesics that Horowitz and Marolf provide is the bounded interval. A correspondence is then supposedly established with reference to the geodesic incompleteness of such a `spacetime' and the failure of the relevant quantum Hamiltonian to be essential self-adjoint. However, incomplete geodesics on the bounded interval fail to exhibit genuine classical singular behaviour precisely because the `spacetime' in question is (trivially) extendible to the unbounded interval, and we thus have incomplete \textit{and extendible} paths. 

This somewhat terminological point aside, it is not at all clear that we should understand the existence of self-adjoint extensions in terms of a `loss of predictability' in correspondence to a classical singularity. When they exist, the family of self-adjoint extensions of the Hamiltonian split the evolution equations into parameterised families of distinct solutions which are, within each group, uniquely defined for all times. Fixing the self-adjoint extension of the Hamiltonian is equivalent to resolving a problem of underdetermination between entire complete solutions based on specification of an additional parameter that is in principle measurable.\footnote{A concrete example of precisely such a situation is provided by experiments that demonstrate the  \emph{Efimov effect} in 3-body quantum systems \cite{efimov:1970,gopalakrishnan:2006,ferlaino:2010}. Here the extension parameter can be measured via a phase shift of incident light on the 3-body system and corresponds to a parameterisation of the breakdown of an effective $1/r^2$ potential. It this context, it is worth noting that such examples suffice to show the invalidity of the general argument of \citeN{earman:2009} that we lack a good physical basis in the choice of self-adjoint extensions and thus only essentially self-adjoint systems display determinate quantum dynamics.} The problematic issue, if there is one, with failure of essential self-adjointness is one uniqueness of solutions rather than existence and thus we have good reason to question formal correspondence to global dynamics classical singularities.  Finally, there is evidence that in the context of Loop Quantum Cosmology models the quantum evolution may in fact be insensitive to the choice of the self-adjoint extension \cite{pawlowski:2012}. This would mean that quantum cosmological evolution may be singular in the sense of SR7 and yet avoid problems with both existence \textit{and} uniqueness. The classical correspondence between SR7 and global dynamics singularity resolution is therefore not well established. At the least the basis under which SR7 can be understood as a necessary condition is not a strong one. 

A further worry regarding SR7 is that explicit models show that \textit{indications} of \textit{local curvature} singular behaviour is found notwithstanding essential self-adjointness. This is a problem since as per classical correspondence we would like our local curvature and global dynamics quantum criteria to coincide as per the classical theory. Consider in particular the model studied by \citeN{gotay:1980}. This analysis focuses on the simplest possible case of a minisuperspace model with classical singularities: a massless scale field with vanishing cosmological constant and positive intrinsic curvature of the constant (cosmological) time hypersurfaces. Gotay and Isenberg perform a reduced phase space quantization based upon choosing the scalar field as the internal clock and constructing a physical Hamiltonian which generate evolution in the relevant clock time. Significantly, within this de-parametrisation scheme the classical singularity is associated with $\phi \rightarrow \infty$. This is because if we write the geometric variable corresponding to the radius of the universe as $R(\phi)$, then we find that $R(\phi)\rightarrow0$ as  $\phi \rightarrow \infty$ \cite{blyth:1975}. It can then be shown that the quantization of the model leads to a dynamics such that  \cite[\S VI]{gotay:1980}:
\begin{equation}
\lim_{\phi\rightarrow\infty}<\psi(\phi)|\hat{R}^2|\psi(\phi)> =0
\end{equation}
This local curvature behaviour obtains despite the fact that the relevant physical Hamiltonian is essentially self-adjoint. There is thus a problem so long as we are guided by the desire to connect local curvature and global dynamics senses, and choose to characterise the former in terms of SR3. One might of course demur with regard to SR3, and insist that the dynamics of the model is locally non-singular since it is generated by a single continuous one-parameter unitary group. Such a response is of course plausible since we ultimately classified SR3 as merely an indicative condition. 

Where does this leave us? The analysis above indicates that SR7 is not reliable as a necessary condition and is at least questionable as a sufficient condition. This motivates us to consider a slightly weaker necessary condition based upon the idea that well-defined dynamics at all times is the relevant physical feature. In particular, let us then consider the possibility of a weaker criterion based on self-adjointness alone, and thus of unitarity itself. In this context, the implication of the example of \cite{gotay:1980}, combined with the indications from SR3, is that unitarity should not be taken as a sufficient condition either. We might, however, consider it to be necessary.\footnote{A weaker, closely related condition would be to treat the vanishing of probability flux into big bang (i.e. zero volume) configurations as a necessary condition for singularity resolution. In this context,  \citeN{demaerel:2019} have recently argued that  vanishing probability flux can be deployed as a necessary \textit{and sufficient} condition. We have excluded detailed consideration of this candidate criterion here on the grounds of interpretational neutrality. Motivation for the sufficient aspect of this candidate criterion appears largely to come from a Bohmian interpretation in which the vanishing probability flux enforces that trajectories do not reach the singularity in finite cosmic proper time. Interpretational neutrality thus puts this approach beyond the remit of our present analysis.} This then leads us to:

\begin{itemize}
\item [\textbf{SR8}:] \textit{Unitarity}. It is a \textit{necessary} condition for \textit{global dynamics} singularity resolution in quantum cosmology that the quantum evolution map is unitary.
\end{itemize}
We would thus have that whilst not all unitary models are globally non-singular, all globally non-singular models must be unitary.

The motivation of SR8 as a global dynamics criterion is directly connected to our earlier discussion of the local curvature criteria SR2 and SR3.\footnote{Further physical motivation can be provided by consideration of a physical model. In particular, \citeN{gotay:1983} consider a  minisuperspace model with a non-unitary quantum Hamiltonian that is associated with a contraction semi-group as opposed to a unitary one-parameter group. Whilst the model does show vanishing expectation values, this `collapse' happens in asymptotic time even when the classical singularity is at finite time. It is thus arguable that the model displays a form of `non-singular' behaviour despite its non-unitarity. One might plausibly insist, however, that the relevant behaviour does not \textit{resolve} the singularity precisely because contractive dynamics is lacking in a clear physical interpretation in the context of cosmology. That said, in the context of approaches to quantum cosmology based upon transition amplitudes there are plausible approaches to making sense of non-unitary dynamics. See \cite{massar:1999}.} In particular, since by definition unitary dynamics in a Hilbert space can never lead to divergent expectation values, we have that unitarity is itself sufficient (although not necessary) for satisfaction of the necessary condition of finite expectation values encoded in the local curvature criterion SR2. The connection with SR3 is more subtle however. On the one hand, there is reasonably strong model-based evidence to support a general conjecture that unitary dynamics can never generate evolution of a non-trivial initial state such that the expectation value of an operator will vanish in finite time \cite[ \S  III.C]{gotay:1983}. On the other hand, the example of  \citeN{gotay:1980} shows that unitary dynamics can lead to locally singular behaviour at the limit of infinite physical clock times.\footnote{Further relevant examples, showing similar singular behaviour with unitary dynamics generated by scalar field clocks, are considered by \citeN[\S VI]{gotay:1983} and \citeN{gielen:2022}.} This then suggests that the possibility that one might supplement unitarity with a further condition regarding the relevant clock time such that the combined global dynamics criterion is both necessary and sufficient.
 
 \subsection{Self-Adjoint Slow Clocks}
 \label{sasc}
 
 The idea of combining particular features of the clock dynamics with unitarity towards a putative criteria for singularity resolution was first suggested by \citeN{gotay:1983} in the context of a conjecture. The conjecture has recently received support in the detailed analysis of \citeN{gielen:2022}, and it will prove well worthwhile to consider it in detail here. Let us first re-formulate the implication of the truth of the conjecture as a criterion and then proceed to defined the relevant terms:
\begin{itemize}
\item [\textbf{SR9}:]  \textit{Self-Adjoint Slow Clocks}. It is \textit{necessary and sufficient} for \textit{global dynamics and local curvature} singularity resolution that the dynamics is generated by a \textit{self-adjoint Hamiltonian} and parameterised by a time variable that it is \textit{dynamically admissible} and \textit{slow}. 
\end{itemize}
There are thus three elements to the criterion. The self-adjointness of the Hamiltonian we have already discussed and is equivalent to unitarity. Above we argued that such a condition is plausibly a necessary but not sufficient condition for global dynamics singularity resolution. Let us then consider the two definitions of the two additional criteria and whether they are suitable to provide a necessary and sufficient condition.
  
Following the definitions of \citeN[p.2604]{gotay:1983}, a time $t$ is \textit{dynamically admissible} if it is \textit{a priori} bounded neither above nor below. The idea is that certain time variables are such that classically they do not range of $(-\infty, +\infty)$ and in such cases self-adjointness of the Hamiltonian leads to strange or perplexing quantum dynamics and should be excluded \cite{gotay:1996}. We will return to re-examine the motivation behind this part of the criterion shortly. We next have that a dynamically admissible time is \textit{fast} if singularities always occurs at $t\rightarrow \pm\infty$ and \textit{slow} if singularities always occurs at $|t|<\infty$ (Gotay and Demaret \citeyear{gotay:1983,gotay:1996}). Fast time dynamics is thus always `complete' and slow time dynamics is always `incomplete' in the relevant clock time. Assuming dynamical admissibility, Gotay and Demaret then present the two conjectures that: F) self-adjoint quantum dynamics in a fast time gauge is always singular; S) Self-adjoint quantum dynamics in a slow time gauge is always non-singular. For our purposes, the significant point is that once if we assume that all times are dynamically admissible and either fast or slow, then we have that the combined truth of S and F leads directly to the necessary and sufficient condition SR9. What is more Gotay and Demaret explicitly frame their conjecture in tandem with SR3, we thus have the enticing possibility that the two may be formally equivalent. Let us assess this final candidate condition according to our methodological principles. 
 
 The first and most obvious worry concerning the condition is that it is claimed within the literature to already be refuted via model based reasoning. In particular, Lemos \citeyear{lemos:1990,lemos:1991,lemos:1996a} claims to have demonstrated the existence of singularities in models that meet the antecedent of the relevant conditional, and thus that the failure of sufficiency. Significantly, the demonstrations rely on the sense of singularity resolution along the lines of SR3 and thus are a direct challenge to the Gotay and Demaret proposal. In response \citeN{gotay:1996} argue that the times used in Lemos's demonstrations are not admissible by their criterion, and thus that the counter examples do not stand. A further response to this response by \citeN{lemos:1996b} argues that, in fact, `the canonical variable [chosen as a clock] is not a priori bounded above or below. It becomes bounded above or below only after selecting one of the two components of the reduced phase space, and thus again unduly restricting the set of allowed initial conditions, which amounts to a mutilation of the original classical model' (p. 5). It if is not entirely clear which side to take in this dispute, then this appears more than anything a product of the vagueness of the Gotay and Demaret dynamical admissibility criterion, in particular, how it should be interpreted in the context of models with discontented reduced phase spaces. At best the role of dynamical admissibility within their candidate criterion is ambiguous, at worst, it could be argued to allow of ad hoc exclusion of possible counter-examples. 
 
This issue brings us to the question of classical correspondence. Whilst the criterion of dynamical admissibility may allow Gotay and Demaret to avoid direct refutation of their conjecture, it also puts considerable stress on the tenability of connection to the relevant classical concepts. This is precisely the worry of Lemos regarding `mutilation of the original classical model'. Moreover, whilst, as we have already argued, it is highly plausible to see vanishing of relevant expectation values as indicative of a quantum analogue of the avoidance of curvature singularities, it is not at all clear why the finiteness of \textit{time variable} in which the singularity is approached should indicate quantum resolution of the singularity. In particular, the idea of slow-clock singularity resolution leads to an odd tension between the classical notion of a global dynamical singularity in terms of geodesic \textit{incompleteness} and the quantum global dynamical notion of the fast-time singular behaviour in terms of the \textit{completeness} of the evolution. Ultimately, much of the physical plausibility of the slow/quick time approach depends upon the connection to the local curvature notion and vanishing of expectation values. If we assume, as seems plausible, that all admissible, self-adjoint, slow-time models are such that the evolution of a non-trivial initial state is never such that the expectation value of an operator will vanish in finite time, then the sufficient part of SR8 seems highly plausible as an indicative condition at the least. 

Next let us consider the question of interpretation neutrality. To the extent to which it has already been argued that reference to expectation values can be understood without reliance on any particular approach to the interpretation of quantum mechanics, the same would hold for Gotay and Demaret's approach. That is, no further quantum interpretational moves are made apart from the understanding of expectation values as tracking physically salient properties. Furthermore, with regard to the problem of time the approach has the attractive feature of being consistent with both internal time and an extrinsic time approaches based on a unimodular time. 

The independence of SR9 with respect to different interpretations of the role of time in quantum cosmology is explicitly demonstrated in the work of Gielen and Men\'{e}ndez-Pidal \citeyear{gielen:2020,gielen:2022}. In particular, in this analysis it is shown that the slow/quick time analysis can be applied to a minisuperspace model with both various internal clock choices and with time chosen as the conjugate momenta to the cosmological constant as per the unimodular approach. This analysis indicates that the unimodular time is automatically an admissible slow time within the self-adjoint quantization of the model and thus this approach can generically be expected to lead to singularity resolution consistent with both SR3 and SR9. This analysis is in full coherence with the earlier treatment of Gryb and Th\'{e}bault \citeyear{Gryb:2017a,Gryb:2017b,Gryb:2018}. By contrast, for the very same model there exist admissible choices of external time in which the self-adjoint quantization leads to fast time evolution and the singularity is not resolved in the sense of SR3. The candidate condition is thus supported in the context of this interpretationally neutral analysis.

This leads us to the perplexing of question of how singularity resolution can depend upon the choice of clock variable in such a robust manner and whether the strength of this connection should lead us to reconsider our stance of interpretational neutrality with regard to the problem of time.\footnote{We should again mention the idea of clock neutrality \cite{hohn:2019,hohn:2020a}. That is, within internal time approaches the expectation that the choice regarding which physical variable is selected as the clock should not have physical significance in the classical or quantum theory. Such choices are understood as the analogue of gauge choices or conventions with regard to spatiotemporal reference frames. On the one hand, there are good formal reasons to expect clock neutrality to obtain in quantum cosmology due to the formal structures of Dirac observables and constraint quantization (H\"{o}hn, Smith, and Lock \citeyear{hohn:2020b,hohn:2021a}). On the other hand, there is direct evidence from the study of quantum mini-superspace models that different clock choice can lead to different dynamics since different clock choices may lead to different kinematical inner products which in turn lead to distinct criteria for unitary dynamics \cite{gielen:2022}. Ultimately the tension between these two approaches is a result of subtle differences in treatment of the quantum analogue of the classical lapse multiplier and the connected issues in the definition of the physical inner product. An interesting discussion of the general implications of these issues discussion can be found in the final part of \cite{gielen:2022b}. See also \cite{malkiewicz:2015,bojowald:2018}.} Given that the connection between clock choice and non-singular behaviour in the quantum theory proves a robust one, and assuming that we have a physical motivation for avoiding singularities in our quantum cosmological model, should we not proscribe against approaches to the problem of time in which the unitary dynamics is with respect to a fast clocks since such clocks, although mathematically well defined, are physically problematic.\footnote{In particular, as noted by \citeN{gielen:2022}, a fast clock can be such that in the clock time the singularity is infinitely far away, but infinity is reached in a finite time. It is rather difficult to give a clear physical interpretation to such phenomenology.} Conversely, might we not use singularity resolution as a motivation for an approach to the problem of time in quantum cosmology in which the clock is guaranteed to be such that unitary evolution is necessarily non-singular. Provision of an answer to such questions warrants a separate detailed analysis that would take us beyond our present remit. For the time being, we will conclude that the self-adjoint slow clock criterion for singularity resolution certainly warrants further formal investigation but is best understood as an indicative condition as things stand.  

\newpage

\section{Summary}

This paper has formulated and analysed nine candidate local curvature and global dynamics criteria for big bang singularity resolution in finite dimensional quantum cosmological models. Our analysis has lead us to endorse a set of four constitutive conditions as detailed below.

\subsubsection*{Necessary Conditions} 

Local Curvature 
\begin{itemize}
\item [\textbf{SR2}:] \textit{Finite Expectation Values}. Given a classical theory in which some class of classical observable diverges it is \textit{necessary} for \textit{local curvature} singularity resolution that the expectation values of the corresponding quantum operators are always finite, if they are ever finite. 
\end{itemize}

\noindent Global Dynamics
\begin{itemize}
\item [\textbf{SR8}:] \textit{Unitarity}. It is a \textit{necessary} condition for \textit{global dynamics} singularity resolution in quantum cosmology that the quantum evolution map is unitary.
\end{itemize}

\subsubsection*{Sufficient Condition} 

Local Curvature
\begin{itemize}
\item [\textbf{SR1}:] \textit{Bounded and Self-Adjoint Physical Operators}. Given a classical quantity that diverges at the singularity, the existence of a quantum representation of this quantity as a bounded and self-adjoint operator on the physical Hilbert space is a \textit{sufficient} condition for \textit{local curvature} singularity resolution.   
\end{itemize}

\subsubsection*{Necessary and Sufficient Condition} 

Global Dynamics 
\begin{itemize}
\item [\textbf{SR6}:] \textit{Quantum Hyperbolicity}. It is a  \textit{necessary} and \textit{sufficient} condition for \textit{global dynamics} singularity resolution that the quantum state should have unique and well-posed evolution equations across or around all sub-manifolds of classically singular configurations.
\end{itemize}

\section*{Acknowledgements}

I am extremely grateful to Abhay Ashtekar, Martin Bojowald, Juliusz Doboszewski,  Erik Curiel, Steffen Gielen, Sean Gryb, Nick Huggett, Mike Schneider, David Sloan, Chris Smeenk, and Jim Weatherall for comments on a draft manuscript, to Henrique Gomes, and Will Wolf for discussion, and to  audience members in Cambridge for valuable feedback on a talk based on this material. Work on this paper was supported by the British Academy through a Mid-Career Fellowship. 

\linespread{1}

\bibliographystyle{chicago}
\bibliography{Masterbib,Masterbib2,singularbib}



\end{document}